\documentclass[epj,referee]{svjour}

\usepackage{amssymb}
\usepackage{graphicx}

\begin{document}

\title{Predicting Missing Links via Local Information}

\author{Tao Zhou\inst{1,2,3,a} \and Linyuan L\"{u}\inst{3} \and Yi-Cheng Zhang\inst{1,2,3}}

\institute {Research Center for Complex System Science, University
of Shanghai for Science and Technology, Shanghai 200093, China \and
Department of Modern Physics, University of Science and Technology
of China, Hefei Anhui 230026, China \and Department of Physics,
University of Fribourg, Chemin du Mus\'ee 3, Fribourg CH-1700,
Switzerland}

\mail{zhutou@ustc.edu}

\abstract{Missing link prediction of networks is of both theoretical
interest and practical significance in modern science. In this
paper, we empirically investigate a simple framework of link
prediction on the basis of node similarity. We compare nine
well-known local similarity measures on six real networks. The
results indicate that the simplest measure, namely Common Neighbors,
has the best overall performance, and the Adamic-Adar index performs
the second best. A new similarity measure, motivated by the resource
allocation process taking place on networks, is proposed and shown
to have higher prediction accuracy than common neighbors. It is
found that many links are assigned same scores if only the
information of the nearest neighbors is used. We therefore design
another new measure exploited information of the next nearest
neighbors, which can remarkably enhance the prediction accuracy.
 % end of PACS codes
} %end of abstract
\maketitle

\section{Introduction}
Many social, biological, and information systems can be properly
described as networks with nodes representing individuals or
organizations and edges mimicking the interactions among them. The
study of complex networks has attracted increasing attention and
become a common focus of many branches of science. Many efforts have
been made to understand the evolution of networks
\cite{Albert2002,Dorogovtsev2002}, the relations between topologies
and functions \cite{Newman2003,Boccaletti2006}, and the network
characteristics \cite{Costa2007}. Very recently, a fresh question is
raised \cite{Redner2008}, that is, how to predict missing links of
networks? For some networks, especially the biological networks such
as protein-protein interaction networks, metabolic networks and food
webs, the discovery of links (i.e., interactions) costs much in the
laboratory or the field, and thus the current knowledge of those
networks is substantially incomplete
\cite{Martinez1999,Sprinzak2003}. Instead of blindly checking all
possible interactions, to predict in advance based on the
interactions known already and focus on those links most likely to
exist can sharply reduce the experimental costs if the predictions
are accurate enough. For some others like the friendship networks in
web society \cite{Grabowski,Hu2009}, very likely but not yet
existing links can be suggested to the relevant users as
recommendations of promising friendships, which can help users
finding new friends and thus enhance their loyalties to web sites.

Majority of the previous works on missing link prediction have used
some external information besides the network topology
\cite{Getoor2005}. Graven \emph{et al.} \cite{Graven2000} predicted
the semantic relationships of the world wide web with the help of
web content. Popescul and Ungar \cite{Popescul2003} designed a
regression model to predict citations made in scientific literature
based on not only the citation graph, but also the authorship,
journal information and content. Taskar \emph{et al.}
\cite{Taskar2003} applied the relational Markov network algorithm to
predict missing links in a network of web pages and a social
network, in which the well-defined attributes of each node are
exploited. O'Madadhain \emph{et al.} \cite{O'Madadhain2005}
constructed local conditional probability models for link
prediction, based on both structural features and nodes' attributes.
The usage of external information can somewhat enhance the
algorithmic accuracy, however, the content and attribute information
are generally not available, thus the applications of above
algorithms are strongly limited. Goldberg and Roth
\cite{Goldberg2003} exploited the neighborhood cohesiveness property
of small-world networks to assess confidence for individual
protein-protein interactions. Liben-Nowell and Kleinberg
\cite{Liben-Nowell} empirically investigated the similarity-based
prediction algorithms for large scientific collaboration networks.
Clauset \emph{et al.} \cite{Clauset2008} designed a prediction
algorithm based on the inherent hierarchical organization of social
and biological networks.

These mentioned works are practically successful in dealing with
specific networks, however, thus far, a comprehensive picture about
the dependence of algorithmic performance on network topology is
lacking. The reason is twofold: (i) the works from engineering and
biological communities have not yet caught up with the current state
of development in characterizing the topologies of complex networks
\cite{Costa2007}, while (ii) the physics community has not paid
enough attention to the link prediction problem. Accordingly, dozens
of important issues are still less explored. For example, one may
concern about how to choose a suitable algorithm given some
structural descriptions of a network, such as small-world phenomenon
\cite{Watts1998}, degree heterogeneity \cite{Barabasi1999}, mixing
pattern \cite{Newman2002}, community structure \cite{Girvan2002},
and so on. In the opposite viewpoint, comparison of the performances
of some prediction algorithms may reveal some structural information
of the networks. It is just like the community structure has
significantly effect on the network synchronizability
\cite{Zhou2007}, while the synchronizing process can be properly
used to reveal the underlying community structure \cite{Arenas2006}.
In addition, the algorithms only based on local information are
generally fast but of lower accuracy, while the ones making use of
the knowledge of global topology are of higher accuracy yet higher
computational complexity \cite{Liben-Nowell}. Can we find a good
tradeoff that provides high quality predictions while requires light
computation?

In this paper, we empirically investigate a simple framework of link
prediction on the basis of node similarity. Although the framework
is simple, it opens a rich space for exploration since the design of
similarity measures is challenging and can be related to very
complicated physics dynamics and mathematical theory, such as random
walk \cite{Gobel1974} and counting problem of spanning trees
\cite{Chebotarev1997}. Here we concentrate on
local-information-based similarities. We compare nine well-known
local measures on six real networks, and the results indicate that
the simplest measure, namely \emph{common neighbors}, has the best
overall performance, which is in accordance with the empirical
results reported in Ref. \cite{Liben-Nowell}. Motivated by the
resource allocation process in transportation networks, we next
propose a new similarity measure, which performs obviously better
than common neighbors, while requires no more information and
computational time. Furthermore, it is found that many links get
same scores under local similarity measures, just like the
degeneracy of energy level. We therefore design a new measure using
the information of the next nearest neighbors, which can break the
``degenerate states" thus remarkably enhance the algorithmic
accuracy. Finally, we outline some future interests in this
direction.

\section{Method}
Considering an undirected simple network $G(V,E)$, where $V$ is the
set of nodes and $E$ is the set of links. The multiple links and
self-connections are not allowed. For each pair of nodes, $x,y\in
V$, every algorithm referred in this paper assigns a score as
$s_{xy}$. This score can be viewed as a measure of similarity
between nodes $x$ and $y$, hereinafter, we do not distinguish
\emph{similarity} and \emph{score}. All the nonexistent links are
sorted in decreasing order according to their scores, and the links
in the top are most likely to exist.

To test the algorithmic accuracy, the observed links, $E$, is
randomly divided into two parts: the training set, $E^T$, is treated
as known information, while the probe set, $E^P$, is used for
testing and no information in probe set is allowed to be used for
prediction. Clearly, $E=E^T\cup E^P$ and $E^T\cap E^P=\varnothing$.
In this paper, the training set always contains 90\% of links, and
the remaining 10\% of links constitute the probe set. We use a
standard metric, area under the receiver operating characteristic
(ROC) curve \cite{Hanely1982}, to quantify the accuracy of
prediction algorithms. In the present case, this metric can be
interpreted as the probability that a randomly chosen missing link
(a link in $E^P$) is given a higher score than a randomly chosen
nonexistent link (a link in $U\setminus E$, where $U$ denotes the
universal set). In the implementation, among $n$ times of
independent comparisons, if there are $n'$ times the missing link
having higher score and $n''$ times the missing link and nonexistent
link having the same score, we define the \emph{accuracy} as:
\begin{equation}
\texttt{AUC}=\frac{n'+0.5n''}{n}.
\end{equation}
If all the scores are generated from an independent and identical
distribution, the accuracy should be about 0.5. Therefore, the
degree to which the accuracy exceeds 0.5 indicates how much better
the algorithm performs than pure chance.

\section{Data}

In this paper, we consider six representative networks drawn from
disparate field: (i) PPI.--- A protein-protein interaction network
containing 2617 proteins and 11855 interactions \cite{Mering2002}.
Although this network is not well connected (it contains 92
components), most of nodes belong to the giant component, whose size
is 2375. (ii) NS.--- A network of coauthorships between scientists
who are themselves publishing on the topic of networks
\cite{Newman2006}. The network contains 1589 scientists, and 128 of
which are isolated. Here we do not consider those isolated nodes.
The connectivity of NS is not good, actually, NS is consisted of 268
connected components, and the size of the largest connected
component is only 379. (iii) Grid.--- An electrical power grid of
western US \cite{Watts1998}, with nodes representing generators,
transformers and substations, and edges corresponding to the high
voltage transmission lines between them. (iv) PB.--- A network of
the US political blogs \cite{Ackland2005}. The original links are
directed, here we treat them as undirected links. (v) INT.--- The
router-level topology of the Internet, which is collected by the
\emph{Rocketfuel Project} \cite{Spring2004}. (vi) USAir.--- the
network of US air transportation system, which contains 332 airports
and 2126 airlines \cite{Pajek}.

\begin{table}
\caption{The basic topological features of six example networks. $N$
and $M$ are the total numbers of nodes and links, respectively.
$N_c$ denotes the size of the giant component, for example, the
entry 2375/92 in the first line means that the network has 92
components and the giant component consists of 2375 nodes. $e$ is
the network efficiency \cite{Latora2001}, defined as
$e=\frac{2}{N(N-1)}\sum_{x,y\in V, x\neq y}d_{xy}^{-1}$, where
$d_{xy}$ is the shortest distance between $x$ and $y$, and
$d_{xy}=+\infty$ if $x$ and $y$ are in two different components. $C$
and $r$ are clustering coefficient \cite{Watts1998} and assortative
coefficient \cite{Newman2002}, respectively. Nodes with degree 1 are
excluded from the calculation of clustering coefficient. $H$ is the
degree heterogeneity, defined as $H=\frac{\langle
k^2\rangle}{\langle k\rangle^2}$, where $\langle k\rangle$ denotes
the average degree.}
\begin{center}
\begin{tabular} {cccccccc}
  \hline \hline
   Nets     & $N$  &  $M$  &  $N_c$ & $e$ & $C$ & $r$ & $H$ \\
   \hline
   PPI & 2617 & 11855 & 2375/92 & 0.180 & 0.387 & 0.461 & 3.73 \\
   NS & 1461 & 2742 & 379/268 & 0.016 & 0.878 & 0.462 & 1.85 \\
   Grid & 4941 & 6594 & 4941/1 & 0.063 & 0.107 & 0.003 & 1.45 \\
   PB & 1224 & 19090 & 1222/2 & 0.397 & 0.361 & -0.079 & 3.13 \\
   INT & 5022 & 6258 & 5022/1 & 0.167 & 0.033 & -0.138 & 5.50 \\
   USAir & 332 & 2126 & 332/1 & 0.406 & 0.749 & -0.208 & 3.46 \\
   \hline \hline
    \end{tabular}
\end{center}
\end{table}

Table 1 summarizes the basic topological features of those networks.
Brief definitions of the monitored topological measures can be found
in the table caption, for more details, please see the review
articles
\cite{Albert2002,Dorogovtsev2002,Newman2003,Boccaletti2006,Costa2007}.
We here give a few remarks for the numbers which may be unexpected
to some readers: (i) It is well known that in the protein-protein
interaction networks, links between highly connected proteins are
systematically suppressed, while those between highly-connected and
low-connected pairs are favored \cite{Maslov2002}. That is to say,
the assortative coefficient should be negative for PPI (for example,
as reported in Ref. \cite{Newman2002}, the Yeast PPI network has an
assortative coefficient -0.156), however, in the present network,
the assortative coefficient is very positive, as 0.461. It is
because the data set used here \cite{Mering2002} is determined from
functional interactions and not from physical interactions. More
detailed discussion can be found in Ref. \cite{Schmith2005}. (ii)
The extremely large clustering coefficient of NS dues to the
specific constructing rule of collaboration networks, namely all the
participants in an act are fully connected. Relevant discussion can
be found in \emph{Appendix B} of Ref. \cite{Zhou2007b}.

\section{Comparison of Nine Similarity Measures Based on Local Information}

\begin{table}
\caption{Accuracies of algorithms, measured by the area under the
ROC curve. Each number is obtained by averaging over 10
implementations with independently random divisions of testing set
and probe set. The abbreviations, CN, Salton, Jaccard, S{\o}rensen,
HPI, HDI, LHN, PA, and AA, stand for Common Neighbors, Salton Index,
Jaccard Index, S{\o}rensen Index, Hub Promoted Index, Hub Depressed
Index, Leicht-Holme-Newman Index, Preferential Attachment, and
Adamic-Adar Index, respectively. The entries corresponding to the
highest accuracies among these nine measures are emphasized by
black. RA and LP are the abbreviations of Resource Allocation Index
and Local Path Index, proposed in Section 5 and Section 6
respectively. The parameter for LP, $\epsilon$, is fixed as
$10^{-3}$.}
\begin{center}
\begin{tabular} {ccccccc}
  \hline \hline
   Measures     & PPI  &  NS  &  Grid & PB & INT & USAir \\
   \hline
CN     & \textbf{0.889}  &  \textbf{0.933}  &  \textbf{0.590} & \textbf{0.925} & \textbf{0.559} & \textbf{0.937} \\
Salton     & 0.869  &  0.911  &  0.585 & 0.874 & 0.552 & 0.898  \\
Jaccard     & 0.888  &  \textbf{0.933}  &  \textbf{0.590} & 0.882 & \textbf{0.559 }& 0.901  \\
S{\o}rensen     & 0.888  &  \textbf{0.933}  &  \textbf{0.590} & 0.881 & \textbf{0.559} & 0.902  \\
HPI     & 0.868  &  0.911  &  0.585 & 0.852 & 0.552 & 0.857  \\
HDI     & 0.888  &  \textbf{0.933}  & \textbf{0.590}  & 0.877 & \textbf{0.559} & 0.895  \\
LHN     & 0.866  &  0.911  &  0.585 & 0.772 & 0.552 & 0.758  \\
PA     & 0.828  &  0.623  &  0.446 & 0.907 & 0.464 & 0.886  \\
AA     & 0.888  &  0.932  &  \textbf{0.590} & 0.922 & \textbf{0.559} & 0.925 \\
\hline
RA     & 0.890  &  0.933  &  0.590 & 0.931 & 0.559 & 0.955 \\
LP     & 0.939  &  0.938  &  0.639 & 0.936 & 0.632 & 0.900 \\
   \hline \hline
\end{tabular}
\end{center}
\end{table}

In this section, we compare prediction accuracies of nine similarity
measures. All these measures are based on the local structural
information contained in the testing set. We first give a brief
introduction of each measure as follows.

(i) \emph{Common Neighbors}.--- For a node $x$, let $\Gamma(x)$
denote the set of neighbors of $x$. In common sense, two nodes, $x$
and $y$, are more likely to have a link if they have many common
neighbors. The simplest measure of this neighborhood overlap is the
directed count, namely
\begin{equation}
s_{xy}=|\Gamma(x)\cap \Gamma(y)|.
\end{equation}

(ii) \emph{Salton Index}.--- Salton index \cite{Salton1983} is
defined as
\begin{equation}
s_{xy}=\frac{|\Gamma(x)\cap \Gamma(y)|}{\sqrt{k(x)\times k(y)}},
\end{equation}
where $k(x)=|\Gamma(x)|$ denotes the degree of $x$. Salton index is
also called cosine similarity in the literature.

(iii) \emph{Jaccard Index}.--- This index was proposed by Jaccard
\cite{Jaccard1901} over a hundred years ago, which is defined as
\begin{equation}
s_{xy}=\frac{|\Gamma(x)\cap \Gamma(y)|}{|\Gamma(x)\cup \Gamma(y)|}.
\end{equation}

(iv) \emph{S{\o}rensen Index}.--- This index is mainly used for
ecological community data \cite{Sorensen1948}, which is defined as
\begin{equation}
s_{xy}=\frac{2\times |\Gamma(x)\cap \Gamma(y)|}{k(x)+k(y)}.
\end{equation}

(v) \emph{Hub Promoted Index}.--- This index is proposed for
quantifying the topological overlap of pairs of substrates in
metabolic networks \cite{Ravasz2002}, defined as
\begin{equation}
s_{xy}=\frac{|\Gamma(x)\cap \Gamma(y)|}{\texttt{min}\{k(x),k(y)\}}.
\end{equation}
Under this measure, the links adjacent to hubs (here, the term
``hub" represents node with very large degree) are probably assigned
high scores since the denominator is determined by the lower degree
only.

(vi) \emph{Hub Depressed Index}.--- Analogously to the above index,
we consider a measure with opposite effect on hubs for comparison,
which is defined as
\begin{equation}
s_{xy}=\frac{|\Gamma(x)\cap \Gamma(y)|}{\texttt{max}\{k(x),k(y)\}}.
\end{equation}

(vii) \emph{Leicht-Holme-Newman Index}.--- This index gives high
similarity to node pairs that have many common neighbors compared
not to the possible maximum, but to the expected number of such
neighbors \cite{Leicht2006}. It is defined as
\begin{equation}
s_{xy}=\frac{|\Gamma(x)\cap \Gamma(y)|}{k(x)\times k(y)},
\end{equation}
where the denominator, $k(x)\times k(y)$, is proportional to the
expected number of common neighbors of nodes $x$ and $y$ in the
\emph{configuration model} \cite{Molloy1995}.

(viii) \emph{Preferential Attachment}.--- The mechanism of
preferential attachment can be used to generate evolving scale-free
networks (i.e., networks with power-law degree distributions), where
the probability that a new link is connected to the node $x$ is
proportional to $k(x)$ \cite{Barabasi1999}. Similar mechanism can
also lead to scale-free networks without growth \cite{Xie2008},
where at each time step, an old link is removed and a new link is
generated. The probability this new link is connecting $x$ and $y$
is proportional to $k(x)\times k(y)$. Motivated by this mechanism, a
corresponding similarity index can be defined as
\begin{equation}
s_{xy}=k(x)\times k(y),
\end{equation}
which has already been suggested as a proximity measure
\cite{Huang2005}, as well as been used to quantify the functional
significance of links subject to various network-based dynamics,
such as percolation \cite{Holme2002}, synchronization \cite{Yin2006}
and transportation \cite{Zhang2007}. Note that, this index requires
less information than all others, namely it does not need to know
the neighborhood of each node. As a consequence, it also has the
minimal computational complexity.

(ix) \emph{Adamic-Adar Index}.--- This index refines the simple
counting of common neighbors by assigning the lower-connected
neighbors more weights \cite{Adamic2003}, which is defined as:
\begin{equation}
s_{xy}=\sum_{z\in \Gamma(x)\cap \Gamma(y)}\frac{1}{\texttt{log}
k(z)}.
\end{equation}

We present the algorithmic accuracies for the six example networks
in Table 2, with those entries corresponding to the highest
accuracies being emphasized by black. To our surprise, the simplest
measure, common neighbors, performs the best. This result is in
accordance with the one reported in Ref. \cite{Liben-Nowell} for
social collaboration networks. Besides CN, Adamic-Adar index
performs the next best for its accuracies are always close to the
best one, while others, such as Jaccard index, S{\o}rensen index and
HDI, perform far worse in the cases for PB and USAir.

Note that, the first seven measures, from CN to LHN, only different
in denominators. If all the nodes have pretty much the same degrees,
corresponding to a very small $H$, then the difference among those
measures becomes insignificant. In addition, for a given network, if
its clustering coefficient is very small, whether two nodes have
common neighbors plays the most important role, while the
denominator is less important. In a word, remarkable difference
among those seven measures can be found only if the monitored
network simultaneously has large clustering coefficient and large
degree heterogeneity, such as PPI, PB and USAir. As shown in Table
2, the performances of those seven algorithms on PB and USAir are
obviously different, but for PPI, they are more or less the same. A
possible reason is that PPI is a very assortative network (i.e.,
$r=0.461$), and thus two nodes of a link tend to have similar
degrees, which reduces the difference in denominators.

The preferential attachment has the worst overall performance,
however, we are interested in it for it requires the minimal
information. One may intuitively think that PA will give good
predictions for assortative networks, while performs badly for
disassortative networks. However, no obvious correlation between
assortative coefficient and algorithmic accuracy based on PA can be
found from our numerical results. The reason is twofold. Firstly,
links between pairs of large-degree nodes contribute positively to
the assortative coefficient and are assigned high scores by PA,
while links between pairs of small-degree nodes also contribute
positively to the assortative coefficient but are disfavored by PA.
Actually, assortative coefficient is an integrated measure involving
many ingredients, and there is no simple relation between this
measure and the performance of PA. Secondly, assortative coefficient
itself is very sensitive to the degree sequence, and a network of
higher degree heterogeneity tends to be disassortative
\cite{ZhouS2007}. Therefore, this single parameter can not reflect
the detailed linking patterns of networks. Clearly, if the
large-degree nodes are very densely connected to each other, and the
small-degree nodes are rarely connected to each other, PA will
perform relatively good. The former relates to the so-called
\emph{rich-club phenomenon} \cite{ZhouS2004}, and we have checked
that PB and USAir exhibit obvious rich-club phenomenon with respect
to their randomized versions (we followed the method proposed by
Colizza \emph{et al.} \cite{Colizza2006}, and they have already
demonstrated the presence of rich-club phenomenon in the air
transportation network). In addition, in USAir, more than 40\% of
nodes are very small local airports, with degrees no more than 3. A
local airport usually connects to a nearby central airport and a
very few hubs, the direct links between two local airports are
rarely found. This topological feature is also favored by PA. As
shown in Table 2, PA gives relatively good predictions for PB and
USAir, in accordance with the above discussion. Note that, all the
other eight measures will automatically assign zero score to the
pair of nodes located in different components. Therefore, PA
performs badly when the network is consisted of many components.
This is the very reason why PA gives very bad predictions for NS,
although NS has clear rich-club phenomenon. We also note that, PA
performs even worse than pure chance for the Internet at router
level and the power grid. In these two networks, the nodes have
well-defined positions and the links are physical lines. Actually,
geography plays a significant role and the links with very long
geographical distances are rare (the empirical analysis of spatial
dependence of links in the Internet can be found in Ref.
\cite{Yook2002}, and the absence of clustering-degree correlation in
the router-level Internet and power grid can be considered as an
indicator of a strong geographical constraint \cite{Ravasz2003}). PA
can not take into account the effect of geographical localization at
all. As local centers, the large-degree nodes have longer
geographical distances to each other than average, correspondingly,
they also have less probability to directly connect to each other.
Actually, these two networks exhibit the anti-rich-club phenomenon,
that is, the link density among very-large-degree nodes are even
lower than the randomized versions. This anti-rich-club effect leads
to the bad performance of PA. In contrast, although USAir has
well-defined geographical positions of nodes, its links are not
physical. Empirical data demonstrated that the air transportation
networks show an inverse relation between clustering coefficient and
degree \cite{Liu2007}, and the number of airline flights is not
sensitive to the geographical distance within the range of about
2000 kilometers \cite{Gastner2006}. As a final remark, comparing Eq.
(8) and Eq. (9), LHN is, to some extent, inverse to PA, therefore
when PA performs badly, LHN will give relatively good predictions,
and vice versa.

\section{Similarity Measure Based on Resource Allocation}

Except PA, all the others introduced in the last section are
neighborhood-based measures. Although they are simple and
mathematically graceful, they are not tightly related to any
physical processes. In this section, motivated by the resource
allocation process taking place in networks \cite{Ou2007}, we
propose a new similarity measure, which has overall higher accuracy
than all the measures mentioned in Section 4.

Considering a pair of nodes, $x$ and $y$, which are not directly
connected. The node $x$ can send some resource to $y$, with their
common neighbors playing the role of transmitters. In the simplest
case, we assume that each transmitter has a unit of resource, and
will averagely distribute it to all its neighbors. The similarity
between $x$ and $y$ can be defined as the amount of resource $y$
received from $x$, which is:
\begin{equation}
s_{xy}=\sum_{z\in \Gamma(x)\cap \Gamma(y)}\frac{1}{k(z)}.
\end{equation}
Clearly, this measure is symmetry, namely $s_{xy}=s_{yx}$.

The algorithmic accuracies on the six example networks are presented
in Table 2, with RA the abbreviation of resource allocation.
Compared with all the nine measures introduced in Section 4, RA
performs the best, especially for the networks (i.e., PB and USAir)
with large clustering coefficient, high degree heterogeneity and
absence of strongly assortative linking pattern. It is observed
that, RA exhibits particularly good performance on USAir. The reason
may be that the resource allocation process is originally proposed
to explain the nonlinear correlation between transportation capacity
and connectivity of each airport \cite{Liu2007,Li2004,Barrat2004}.

Note that, although being resulted from different motivations
\cite{Adamic2003,Ou2007}, the Adamic-Adar index and resource
allocation index have very similar form, indeed, they both depress
the contributions of common neighbors with high degrees. The
difference between $\frac{1}{\texttt{log}k(z)}$ and $\frac{1}{k(z)}$
(see Eq. (10) and Eq. (11)) is insignificant if the degree, $k(z)$,
is small, while it is great if $k(z)$ is large. Therefore, when the
average degree is very small, the prediction results of AA and RA
are very close, while for the networks of large average degree, such
as PB and USAir, the results are clearly different and the RA
performs better, which implies that the punishment on large-degree
common neighbors of AA is insufficient.

RA can be extended to the asymmetry case. Assuming a unit of
resource is located in $x$, which will be equally send to all $x$'s
neighbors, each of which will equally distribute the received
resource one step further to all its neighbors. The amount of
resource a node $y$ received can be considered as the importance of
$y$ in $x$'s sense, denoted as
\begin{equation}
s_{xy}=\frac{1}{k(x)}\sum_{z\in \Gamma(x)\cap
\Gamma(y)}\frac{1}{k(z)}.
\end{equation}
In this case, $s_{xy}\neq s_{yx}$. This idea has already found its
applications in a personalized recommendation algorithm of bipartite
user-object networks \cite{Zhou2007c,Zhou2008}.

\section{Improving Algorithmic Accuracy by Breaking the Degenerate States}

The neighborhood-based measures require only the information of the
nearest neighbors, therefore have very low computational complexity.
However, the information usually seems insufficient, and the
probability that two node pairs are assigned the same score is high.
That is to say, the neighborhood-based similarity is less
distinguishable from each other. If we consider the score assigned
to a node pair as its energy, then many node pairs crowd into a very
few energy levels. Taking INT as an example, there are more than
$10^7$ node pairs, 99.59\% of which are assigned zero score by CN.
For all the node pairs having scores higher than 0, 91.11\% of which
are assigned score 1, and 4.48\% are assigned score 2. Using a
little bit more information involving the next nearest neighbors may
break the ``degenerate states" and make the scores more
distinguishable. Denoting $A$ the adjacent matrix, where $A_{xy}=1$
if $x$ and $y$ are directly connected, and $A_{xy}=0$ otherwise.
Obviously, $(A^2)_{xy}$ is the number of common neighbors of nodes
$x$ and $y$, which is also equal to the number of different paths
with length 2 connecting $x$ and $y$. And if $x$ and $y$ are not
directly connected (this is the case we are interested in),
$(A^3)_{xy}$ is equal to the number of different paths with length 3
connecting $x$ and $y$. The information contained in $A^3$ can be
used to break the degenerate states, and thus we define a new
measure as
\begin{equation}
S=A^2+\epsilon A^3,
\end{equation}
where $S$ denotes the similarity matrix and $\epsilon$ is a free
parameter. We call it \emph{Local Path} (LP) index for it makes use
of the information of local paths with lengths 2 and 3. Clearly, LP
degenerates to CN when $\epsilon=0$. Here, the information in $A^3$
is only used to break the degenerate states, therefore $\epsilon$
should be a very small number close to zero (of course, given a
network, one can tune $\epsilon$ to find its optimal value
corresponding to the highest accuracy, however, this optimal value
is different for different networks, and a parameter-dependent
measure is less practical in dealing with huge-size networks since
the tuning process may take too long time). In the real
implementation, we directly count the number of different paths with
length 3, which is much faster than the matrix multiplication, and
thus Eq. (13) is also based on local calculation.

The algorithmic accuracies on the six example networks are presented
in Table 2, where this measure is denoted by LP and the parameter is
fixed as $\epsilon=10^{-3}$. It is happy to see that the accuracy,
except for USAir, can be largely enhanced by LP. In USAir, the
large-degree nodes are densely connected and share many common
neighbors. Some links among large-degree nodes are removed into the
probe set. Even without the contribution of $\epsilon A^3$, those
links are assigned very high scores, thus the additional item,
$\epsilon A^3$, changes little of their relative positions.
Considering two small local airports, $x$ and $y$, which are
connected to their local central airports, $x'$ and $y'$. Of course,
many hubs are common neighbors of $x'$ and $y'$, and $x'$ and $y'$
may be directly connected. If the link $(x,x')$ is removed, the
similarities between $x$ and other nodes are all zero for both CN
and LP. If $(x,x')$ exists, by LP, the similarities $s_{xy'}$ (by
$x$-$x'$-hub-$y'$), $s_{xy}$ (by $x$-$x'$-$y'$-$y$), and $s_{xh}$
where $h$ represents a hub node (by $x$-$x'$-hub-$h$ and/or
$x$-$x'$-$y'$-$h$) are positive due to the contributions of paths
with length 3. There are many links connecting small local airports
and local centers, some of which are removed, and the others are
kept in the testing set. According to the above discussion, the
removed links have lower score than the nonexistent links due to the
additional item $\epsilon A^3$. In a word, the very specific
structure of USAir (the hierarchical organization consisted of hubs,
local centers and small local airports) makes the LP worse than the
simple CN. In this specific case, we can break the degenerate states
in the opposite direction by setting $\epsilon$ being equal to
$-10^{-3}$, which lead to an accuracy 0.945, higher than the one by
CN, 0.937.

\section{Conclusion and Discussion}

In this paper, we empirically compared some link prediction
algorithms based on node similarities. All the similarity measures
discussed here, including the two newly proposed ones, can be
obtained by local calculations. Numerical results on the nine
well-known measures indicate that: (i) The simplest measure, common
neighbors, performs the best, and the Adamic-Adar index is the
second; (ii) Remarkable difference among these measures, excluding
the Adamic-Adar index and the preferential attachment, can be
observed only if the monitored network is with large clustering
coefficient, high degree heterogeneity, and absence of strongly
assortative linking pattern; (iii) The preferential attachment
performs relatively good if the monitored network has the rich-club
phenomenon.

We proposed a new measure, RA, motivated by the resource allocation
process, which is equivalent to the one-step random walk starting
from the common neighbors. This measure has a similar form to the
Adamic-Adar index, but performs better, especially for the networks
with high average degrees. We guess RA is particularly suitable for
link prediction of transportation networks, whose validity needs
further evidence from more empirical results. We here strongly
recommend this measure to relevant applications and theoretical
analyses, not only for its good performance, but also for its
simplicity and grace.

Furthermore, we found that many links are assigned same scores based
on the local measures using the information of the nearest neighbors
only. Exploitation of some additional information of the next
nearest neighbors can therefore break the degenerate states and
enhance the algorithmic accuracy. In real applications, the
algorithms based on global calculations may be less efficient for
they require long time and/or huge memory, while the algorithms only
exploited very local information may be less effective for their low
accuracies. A properly designed algorithm can provide a good
tradeoff just like the LP index presented in this paper. Indeed, it
is shown recently that the LP index provides competitively accurate
predictions compared with the indices making use of global
information \cite{Lu2009}. A similar idea has also been adopted in
the network-based traffic dynamics, where the information of the
next nearest neighbors can sharply enhance the traffic efficiency
compared with the case in which only the information of the nearest
neighbors is known \cite{Tadic2004}.

Although the framework adopted here is very simple, it opens a rich
space for investigation since in principle, all algorithms can be
embedded into this framework differing only in the similarity
measures. Besides ones discussed in this paper, a number of
similarities are based on the global structural information, such as
the average commute time of random walk \cite{Gobel1974}, the number
of spanning trees embedding a given node pair \cite{Chebotarev1997},
the pseudoinverse of the Laplacian matrix \cite{Fouss2007}, and so
on. Some other similarity measures are even more complicated,
depending on parameters. These include the Katz index
\cite{Katz1953} and its variant \cite{Leicht2006}, the transferring
similarity \cite{Sun2008}, the PageRank index \cite{Brin1998}, and
so on. These measures may give better predictions than the local
ones, however, the calculation of such measures, including
determination of the optimal parameters for specific networks, is of
high complexity, and thus infeasible for huge-size networks. Anyway,
up to now, we lack systematic comparison and clear understanding of
the performances of these measures, which is set as our future
works.

Empirical analysis on more real networks as well as more known and
newly proposed similarity measures is very valuable for building up
knowledge and experience, and we can expect a clear picture of this
issue can be completed by putting together of many fragments from
respective empirical studies. However, the empirical results may be
not clear at all since many unknown and uncontrollable ingredients
are always mixed together in real networks. An alternative way is to
build artificial network models with controllable topological
features, and to compare the prediction algorithms on these models
(see Ref. \cite{Lu2009} about the comparison of link prediction
algorithms on modeled networks with controllable density and noise
strength).

\begin{acknowledgement}

This work is partially supported by the Swiss National Science
Foundation (Project 205120-113842) and Physics of Risk through
project C05.0148. T.Z. acknowledges the National Natural Science
Foundation of China (Grant Nos. 10635040 and 60744003). L.L.
acknowledges the National Scholarship Fund of China Scholarship
Council.

\end{acknowledgement}

\end{document}